\documentclass[%
 aip,
 amsmath,amssymb,
 reprint,%
]{revtex4-1}

\usepackage{graphicx}
\usepackage{xcolor}
\usepackage{dcolumn}
\usepackage{bm}
\usepackage{verbatim}
\usepackage{url}

\usepackage[utf8]{inputenc}
\usepackage[T1]{fontenc}
\usepackage{mathptmx}
\usepackage{etoolbox}

\makeatletter
\def\@email#1#2{%
 \endgroup
 \patchcmd{\titleblock@produce}
  {\frontmatter@RRAPformat}
  {\frontmatter@RRAPformat{\produce@RRAP{*#1\href{mailto:#2}{#2}}}\frontmatter@RRAPformat}
  {}{}
}%
\makeatother
\begin{document}

\preprint{AIP/123-QED}

\title[Calibration of scintillator-based X-ray detectors for broadband laser-driven X-ray radiation]{Calibration of scintillator-based X-ray detectors for broadband laser-driven X-ray radiation\\}
\author{Orsolya Morvai}
\thanks{These authors contributed equally to this work.}
\affiliation{Extreme Light Infrastructure ERIC, ELI Beamlines Facility, Za Radnicí 835, 252 41 Dolní Břežany, Czechia}
\affiliation{Czech Technical University in Prague, Faculty of Nuclear Sciences and Physical Engineering, Břehová 7, 115 19 Prague 1, Czechia}
\email[Corresponding author: ]{orsolya.morvai@eli-laser.eu}

\author{Benoit Lefebvre}
\thanks{These authors contributed equally to this work.}
\affiliation{Extreme Light Infrastructure ERIC, ELI Beamlines Facility, Za Radnicí 835, 252 41 Dolní Břežany, Czechia}

\author{Marcel Lamač} \affiliation{Extreme Light Infrastructure ERIC, ELI Beamlines Facility, Za Radnicí 835, 252 41 Dolní Břežany, Czechia}  \author{Uddhab Chaulagain} \affiliation{Extreme Light Infrastructure ERIC, ELI Beamlines Facility, Za Radnicí 835, 252 41 Dolní Břežany, Czechia} \author{Petr Odstrčil} \affiliation{Extreme Light Infrastructure ERIC, ELI Beamlines Facility, Za Radnicí 835, 252 41 Dolní Břežany, Czechia} \author{Dominik Čáp} \affiliation{Extreme Light Infrastructure ERIC, ELI Beamlines Facility, Za Radnicí 835, 252 41 Dolní Břežany, Czechia} \affiliation{Czech Technical University in Prague, Faculty of Nuclear Sciences and Physical Engineering, Břehová 7, 115 19 Prague 1, Czechia} \author{Vojtěch Janota} \affiliation{Extreme Light Infrastructure ERIC, ELI Beamlines Facility, Za Radnicí 835, 252 41 Dolní Břežany, Czechia} \affiliation{Czech Technical University in Prague, Faculty of Nuclear Sciences and Physical Engineering, Břehová 7, 115 19 Prague 1, Czechia} \author{Alfred Haavaan Mishi} \affiliation{Extreme Light Infrastructure ERIC, ELI Beamlines Facility, Za Radnicí 835, 252 41 Dolní Břežany, Czechia} \affiliation{Czech Technical University in Prague, Faculty of Nuclear Sciences and Physical Engineering, Břehová 7, 115 19 Prague 1, Czechia} \author{Romain Caye} \affiliation{Université Paris-Saclay, Institut d'Optique Graduate School, 91120 Palaiseau, France} \author{Jaroslav Nejdl} \affiliation{Extreme Light Infrastructure ERIC, ELI Beamlines Facility, Za Radnicí 835, 252 41 Dolní Břežany, Czechia} \affiliation{Czech Technical University in Prague, Faculty of Nuclear Sciences and Physical Engineering, Břehová 7, 115 19 Prague 1, Czechia} \author{Antonia Morabito} \affiliation{Extreme Light Infrastructure ERIC, ELI Beamlines Facility, Za Radnicí 835, 252 41 Dolní Břežany, Czechia}

\date{\today}

\begin{abstract}
Laser-driven X-ray sources produce broadband radiation with substantial
shot-to-shot fluctuations, requiring calibrated detector-response models
for quantitative measurements of photon fluence and spectral distribution.
Scintillator-based flat-panel detectors, originally developed primarily
for medical and industrial X-ray imaging, are increasingly being adopted
for diagnostics of laser--plasma-based X-ray sources because they provide
large-area, spatially resolved detection. We report the calibration of two
complementary X-ray detector systems: a CsI:Tl-based flat-panel detector
and a plastic-scintillator filter-stack spectrometer intended for spectral
reconstruction. Both detectors were characterized using well-defined
ISO~4037 N-series reference radiation qualities, providing controlled
polychromatic X-ray fields for establishing their signal response and
signal-to-fluence conversion.

\end{abstract}

\maketitle


\section{\label{sec:level1}Introduction}

X-ray sources are essential tools in science, medicine, and industry, enabling nondestructive probing of internal structures and ultrafast time-resolved studies~\cite{attwood1999xray,emma2010first,Ullrich2012}. Conventionally, they are produced either by energetic electrons decelerating in solid targets or by relativistic electron beams in synchrotron and free-electron-laser facilities~\cite{AlsNielsen2011}.

Laser–plasma accelerators have recently enabled compact ultrafast X-ray sources. In particular, betatron radiation from laser-wakefield accelerators (LWFAs) has attracted strong interest due to its femtosecond duration, micrometer-scale source size, and broad spectral bandwidth~\cite{rousse2004production,Kneip2010,Corde2013Betatron}, making it promising for phase-contrast imaging, tomography, ultrafast X-ray spectroscopy, and studies of high-energy-density matter~\cite{Wenz2015BetatronPCI,Fourmaux2011SingleShotBetatron}.

The broadband spectra, shot-to-shot fluctuations, and increasing photon energies of LWFA betatron sources place demanding requirements on X-ray diagnostics, as no single detector technology is optimal across this full parameter space.

Semiconductor-based direct detectors (Si, CdTe, CZT) offer high spatial resolution and fast response but suffer efficiency losses at higher energies due to finite detector thickness and charge-collection effects~\cite{Knoll2010, Mozzanica2012,Pickford2020EffectsThickness}. Scintillator-based indirect detectors provide high efficiency and large-area coverage, but their response depends on a complex chain involving scintillation yield, optical transport, and readout efficiency~\cite{Sellin2007,Wood2016Calibration}.

For spectral measurements of intense pulsed sources, filter stack spectrometers (FSSs) provide a complementary approach by reconstructing the incident spectrum from depth-dependent dose deposition in layered detector-filter assemblies, which can be tailored to specific energy ranges~\cite{Armstrong2023,laso_garcia_2022_calorimeter,Lefebvre2024,fauvel_2025_compact_gamma_spectrometer}.

Reliable characterization of LWFA-based X-ray sources therefore requires detectors with well-understood energy-dependent response functions, making accurate calibration essential for quantitative fluence and spectral reconstruction.

In this work, we calibrate and model two complementary detector systems using ISO~4037-1 reference X-ray fields~\cite{ISO4037-1:2019}. The first is a Varex Imaging XRD 0822 flat-panel detector~\cite{varex2022xrd0822}, a CsI:Tl-based indirect imaging system. The second is a custom plastic-scintillator-based FSS read out by a CMOS camera (Section~\ref{sec:calib-fss}). Both systems are modeled and validated through simulation and experiment.

The detectors are intended for future experiments at the L3-Gammatron beamline~\cite{Gammatron,chaulagain2025bright,photonics9110853} of the Extreme Light Infrastructure European Research Infrastructure Consortium (ELI ERIC), where the flat-panel detector will provide imaging and filter-based spectral information, while the FSS will enable independent fluence and spectrum reconstruction.

This article is organized as follows: Section \ref{sec:calib-methology} describes the experimental setup used to generate the reference X-ray fields and calibrate both detector systems. Section ~\ref{sec:calib-panel} presents the response model of the flat-panel detector, while Section ~\ref{sec:calib-fss} details the response model of the custom FSS. Both detector response models are validated against the reference radiation fields in their respective sections. Section \ref{sec:conclusion} summarizes the main findings and conclusions of this work.

\begin{figure} [htb!]
    \centering
\includegraphics[width=0.49\textwidth]{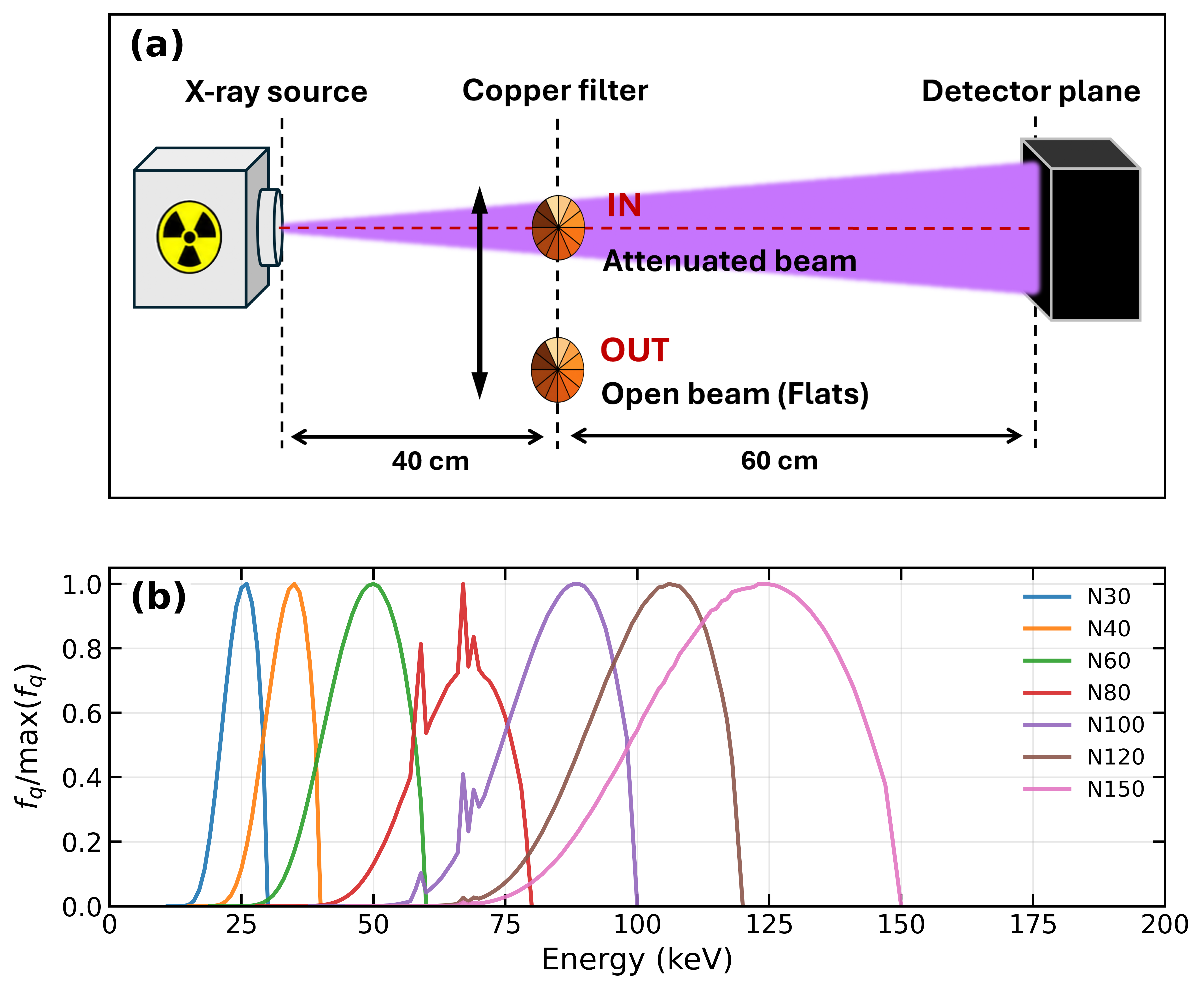}
    \caption{
    \textbf{(a)} Calibration setup. The Varex flat-panel detector (FP) and the in-house filter-stack spectrometer (FSS) were placed sequentially at the detector plane and irradiated using the same source geometry. For FP calibration, attenuated images were acquired with a copper filter inserted into the beam path, while open-beam flat-field images were acquired without the filter. The FSS measurements were performed only without the copper filter wheel. \textbf{(b)} Spectra of the calibration X-ray fields at the detector plane without copper attenuation, normalized to their respective maxima.}
    \label{fig:experimentalsetup}
\end{figure}

\section{Calibration methodology}
\label{sec:calib-methology}

The calibration measurements were performed at the National Radiation Protection Institute (SÚRO) in Prague, Czech Republic~\cite{SURO_Prague}. The facility is equipped with an X-ray source compliant with ISO~4037-1,
which specifies the tube voltages and corresponding filtrations
(Be, Al, or Cu) required to generate a broad range of standardized
narrow-spectrum X-ray qualities with approximately Gaussian spectral
distributions.

For the present study, the N-series radiation qualities were selected, as they provide a good compromise between beam intensity and spectral width. A tungsten-anode X-ray tube was operated at tube voltages ranging from 30 to 150 kV, producing spectrally distinct reference beam qualities relevant to current and emerging LWFA X-ray sources \cite{rousse2004production,Corde2013Betatron}. In particular, this energy range is appropriate for the CsI:Tl scintillator layer of the Varex flat-panel detector. The X-ray qualities are designated using the notation NXXX, where XXX denotes the X-ray tube voltage in kilovolts.
During calibration, the detectors were positioned at a distance of precisely 100~cm from the X-ray source with the detector front surface centered on the X-ray beam axis, as shown in Figure \ref{fig:experimentalsetup}(a).

For each X-ray quality, the rate of kinetic energy released per unit mass  (kerma) of air, in units of $\mathrm{Gy\,s^{-1}}$ was measured beforehand using a calibrated air-ionization chamber placed at the same location. To determine the photon fluence rates corresponding to the measured air kerma rates, the normalized spectral distributions, denoted $f_{q}(E)$ where $q$ denotes the quality, were first calculated using the SpekCalc software package~\cite{poludniowski2007calculation1,poludniowski2007calculation2,poludniowski2009spekcalc,Andreo2012, Shikhaliev_2011} based on the specified beam qualities and source-to-detector distance. The calculated spectra are shown in Fig.\ref{fig:experimentalsetup}(b).

The integrated photon fluence rate, $\dot{\Phi}$, was then obtained from the following relation:
\begin{equation} 
\dot{K}_{\mathrm{col}} = \dot{\Phi} \int f_{q}(E)\,  E \left( \frac{\mu_{\mathrm{en}}(E)}{\rho} \right)_{\mathrm{air}}  \, \mathrm{d}E, 
\label{eq:kerma_measurement} 
\end{equation}
\noindent
where $\dot{K}_{\mathrm{col}}$ is the collisional air kerma rate, $\dot{\Phi}$ is the integrated photon fluence rate, $E$ denotes the photon energy, and $\left(\mu_{\mathrm{en}}(E)/\rho\right)_{\mathrm{air}}$ is the energy-dependent mass energy-absorption coefficient of air~\cite{ICRU85}. The values of the mass energy-absorption coefficient were taken from the NIST Standard Reference Database 126~\cite{HubbellSeltzerNIST126}.
\noindent
The X-ray qualities employed in this study, together with their measured air kerma rates $\dot{K}_{\mathrm{col}}$ and corresponding calculated photon fluence rates $\dot{\Phi}$, are summarized in Table~\ref{tab:xrayquality}.

\begin{table}[htb!]
\centering
\begin{tabular}{l|l|l}
X-ray quality & $\dot{K}_{\mathrm{col}}$ [{\textmu}Gy/s] & $\dot{\Phi}$ [$\times 10^7$ cm$^{-2}$ s$^{-1}$] \\
\hline
\hline
N30  & 55.4 & 3.93  \\
N40  & 24.2 & 3.42  \\
N60  & 39.3 & 10.46 \\
N80  & 20.9 & 6.83  \\
N100 & 10.0 & 3.05  \\
N120 & 11.2 & 2.92  \\
N150 & 79.8 & 17.36 \\
\hline
\end{tabular} 
\caption{X-ray qualities used in this study, with measured air kerma rates and corresponding calculated photon fluence rates.}
\label{tab:xrayquality}
\end{table}
The X-ray beam fluence was assumed to be uniform over the fiducial surface of both detectors. This assumption was confirmed by independent measurements with small ionization chambers, which showed that the decrease in air kerma rate was less than 1\% within a transverse distance of 10~cm from the beam axis at the measurement position.

\noindent

The standard error of the kerma measurements is 1\%. 
For some of the flat-panel detector measurements, a copper filter wheel
with a radius of 3~cm was positioned on the beam axis, 40~cm from the
X-ray source and 60~cm upstream of the detector, to further modify the
photon spectrum. The wheel consisted of 11 copper sectors with
thicknesses ranging from 17~$\mu$m to 6~mm and one open sector for
open-beam measurements.

\noindent

Secondary electrons generated in the copper filters were neglected, since their range in air is much shorter than the 60 cm filter-to-detector distance for the investigated X-ray energies (up to 150 keV). ~\cite{Berger2005}
\noindent
Measurements acquired through the individual copper sectors provided
additional calibration conditions. Measurements performed without the
filter wheel are hereinafter referred to as flat-field measurements.

\section{Calibration of the flat-panel}
\label{sec:calib-panel}

\subsubsection{Detector description} \label{sec:fp_detector} 

The flat-panel detector used in this work was a Varex Imaging XRD 0822 digital flat-panel detector. It has a pixel matrix of $1024\times1024$ pixels with a pixel pitch of $200~\mu\mathrm{m}$, corresponding to an active area of $204.8\times204.8~\mathrm{mm}^2$.
\newline

\begin{figure} [htb!]
\includegraphics[width=0.49\textwidth]{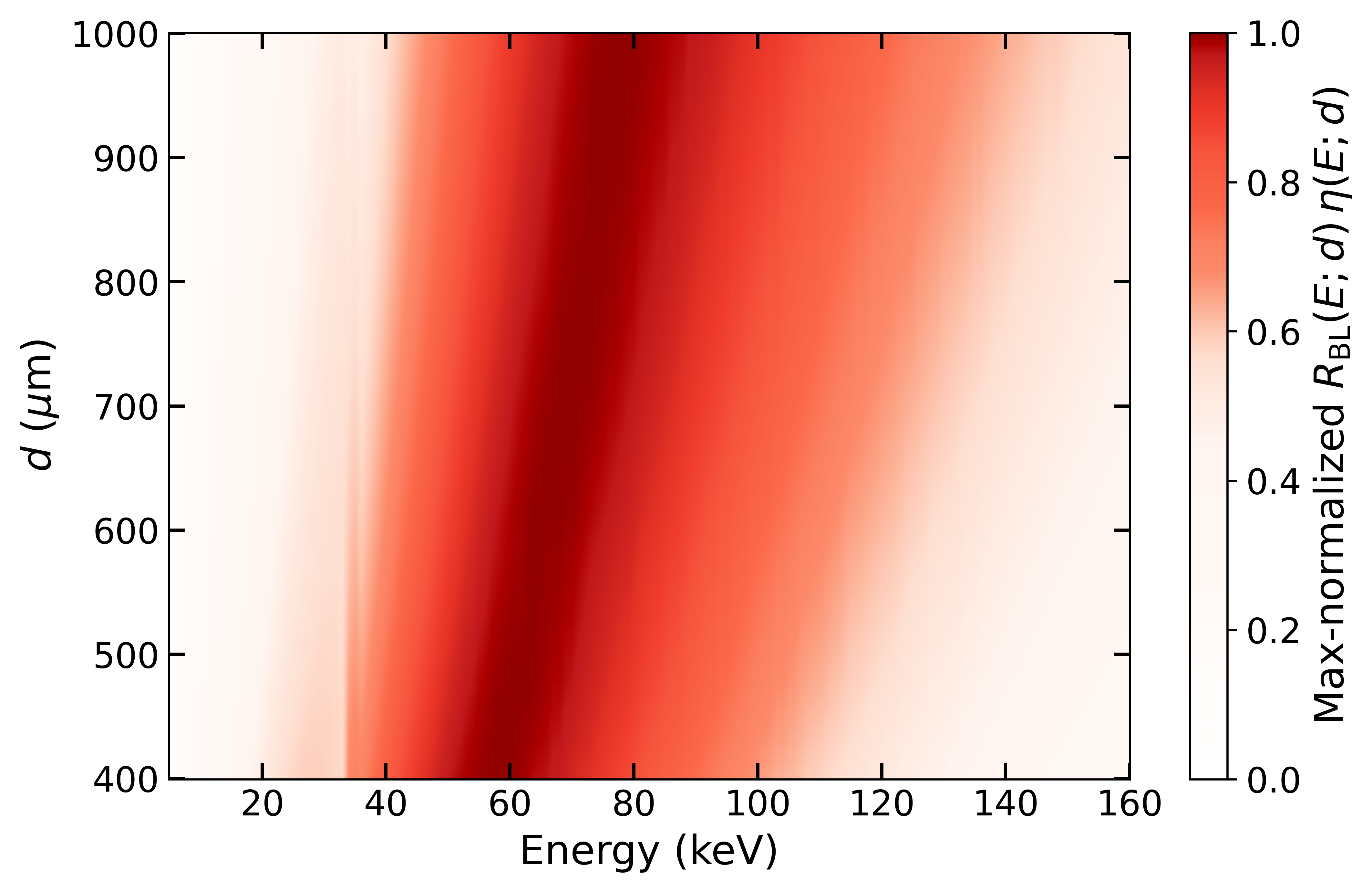}
    \caption{Max-normalized detector response, proportional to $R_{\mathrm{BL}}(E;d)\eta(E;d)$, as a function of photon energy and CsI:Tl scintillator thickness. The response includes Beer--Lambert attenuation in the scintillator and the FLUKA-derived correction for incomplete local energy deposition. FLUKA simulations were performed for discrete scintillator thicknesses and linearly interpolated between simulated values. Each thickness-dependent response curve was normalized to its own maximum to emphasize the spectral response shape.}
    \label{fig:function}
\end{figure}

The detailed composition of the detector stack, including protective entrance layers, optical coupling layers, and the exact CsI:Tl scintillator thickness, is not specified by the manufacturer. The detector was operated using the manufacturer-provided acquisition software. Images were acquired with exposure times varying between $1$ and $10~\mathrm{s}$ depending on the X-ray quality. The exposure time was adjusted to avoid signal saturation.
\newline

The detector readout gain was controlled through a dimensionless software setting. The available gain settings vary between $0.25$ and $8$. All calibration measurements used in the response reconstruction were acquired at gain $1$. We found empirically that the signal scales linearly with the gain setting. Since all calibration data included in the fit were acquired using the same
readout setting, the corresponding signal-conversion factor is absorbed into
the single fitted effective conversion parameter $G$.

\subsubsection{Detector response model}
\label{sec:flat-panel-resp-model}

The flat-panel signal was modeled based on the energy deposited by ionizing radiation in the CsI:Tl scintillator. The deposited energy was converted to the recorded detector signal through
scintillation-light generation, optical collection, photodiode
light-to-charge conversion, and electronic readout. These processes were
represented collectively by the effective conversion factor $G$. Absorption in the CsI:Tl layer was described using Beer--Lambert attenuation. An energy-dependent correction accounting for the energy deposited in the scintillator following photon interaction was derived from Monte Carlo simulation, which is described below.

For an incident photon of energy $E$, we define
$R_{\mathrm{BL}}(E;d)$ as the idealized Beer--Lambert estimate of the
mean energy deposited in a CsI:Tl scintillator of thickness $d$, under
the assumption of full-energy deposition:

\begin{equation}
R_{\mathrm{BL}}(E;d)
=
E\left[
1-\exp\left(-\mu_{\mathrm{CsI:Tl}}(E)d\right)
\right],
\label{eq:rbl}
\end{equation}

\noindent

where $\mu_{\mathrm{CsI:Tl}}(E)$ is the energy-dependent linear attenuation coefficient of CsI:Tl~\cite{lambert1760photometria}. The ratio $R_{\mathrm{BL}}(E;d)/E$ gives the probability that an incident photon is removed from the uncollided primary beam. Equation~\eqref{eq:rbl} assumes that each such photon deposits its full incident energy in the scintillator. Since the actual CsI:Tl thickness of the panel is not specified by the manufacturer, $d$ was treated as a free parameter representing an effective scintillator thickness.

The detector response per incident photon can be written 

\begin{equation} R(E;d,G) = G\,\eta(E;d)\,R_{\mathrm{BL}}(E;d), \label{eq:response} \end{equation}

\noindent
where $G$ is an effective energy-to-signal conversion factor with units of analog-to-digital units per keV (ADU/keV). 

The energy deposition correction factor $\eta(E,d)$ was obtained from radiation transport simulations performed with the FLUKA Monte Carlo code~\cite{FLUKA2022,BATTISTONI201510}. For this series of simulations and the remainder of this work, version 4.5-1 was used. Moreover, transport and production cuts of 1~keV and 5~keV are used for photons and electrons, respectively. All other default physics settings were used. 

In each simulation, monochromatic photons were generated normally incident on a large CsI:Tl scintillator slab, and the energy deposited in the scintillator per primary photon was scored. The correction factor $\eta$ was defined as the ratio of
the scored deposited energy to the energy deposition predicted by the Beer--Lambert
approximation, which assumes that the full photon energy is absorbed locally
following an interaction. It therefore accounts for incomplete energy
deposition within the scintillator as predicted by the FLUKA
radiation-transport simulation. Simulations were performed for photon energies ranging from 10 to 150 keV in 1 keV increments and for scintillator thicknesses ranging from 400 to 1000 $\mu$m in 100 $\mu$m increments. The statistical uncertainty of the simulated values of $\eta$ was below 1\% for all simulated cases. For the subsequent analysis, the discrete simulation results were used to construct a two-dimensional interpolation function of photon energy $E$ and scintillator thickness $d$. The simulations predict energy losses of up to 20\% over the full parameter-space considered.

The resulting energy-dependent response functions are shown in Fig.~\ref{fig:function}. The curves represent the max-normalized response proportional to $R_{\mathrm{BL}}(E;d)\eta(E;d)$ for different CsI:Tl scintillator thicknesses.

\begin{figure*}[htb!]
    \centering
    \includegraphics[width=\textwidth]{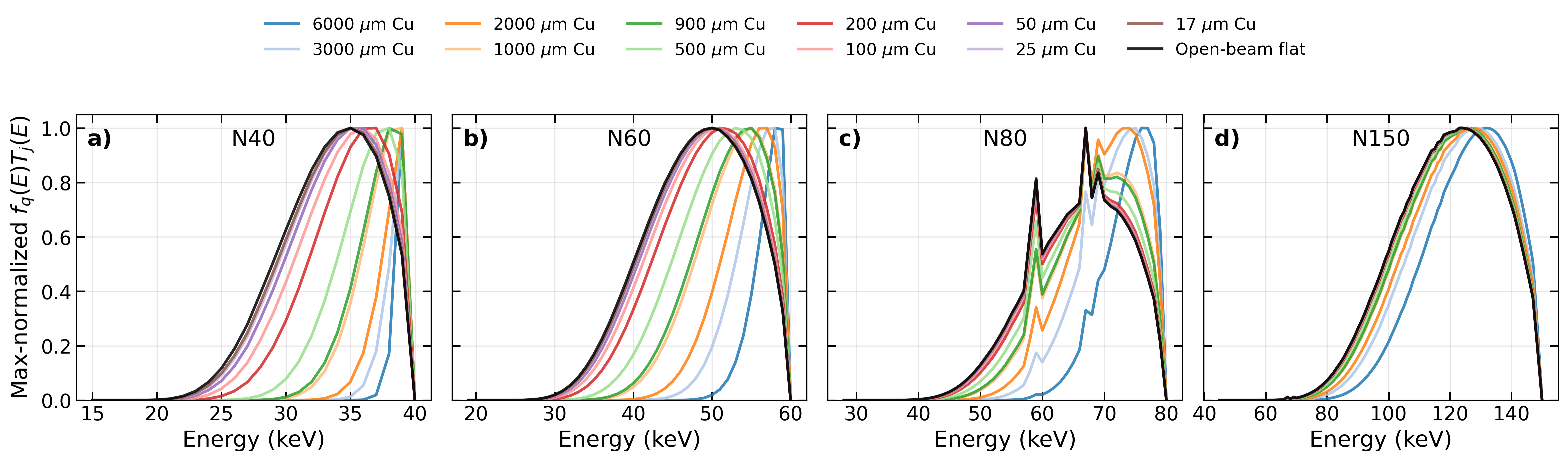}
    \caption{
    Conditioned spectra of the X-ray qualities N40, N60, N80, and N150 after transmission through the copper filters. Increasing copper thickness progressively suppresses the low-energy part of the spectrum, resulting in spectral hardening of the transmitted radiation.
    }
    \label{fig:spectra_filters}
\end{figure*}

Copper attenuators arranged in a filter wheel were used to generate
additional calibration conditions by modifying the incident X-ray field.
For each X-ray quality and filter configuration, the incident spectrum was modified by the corresponding copper transmission and subsequently weighted by the energy-dependent scintillator response.

\noindent

For a copper filter $j$ of thickness $d_{\mathrm{Cu},j}$, the transmission was calculated as

\begin{equation} T_j(E) = \exp\left[-\mu_{\mathrm{Cu}}(E)d_{\mathrm{Cu},j}\right], \label{eq:cu_transmission} \end{equation}

\noindent
where $\mu_{\mathrm{Cu}}(E)$ is the energy-dependent linear attenuation coefficient of copper. The transmitted photon spectra are therefore proportional to $f_q(E)T_j(E)$, as shown for selected beam qualities and filter thicknesses in Fig.~\ref{fig:spectra_filters}.

A calibration condition is denoted by the index $k=(q,j)$, corresponding to one beam-quality/filter pair. For a pixel area $A_{\mathrm{pix}}$, exposure time $t_{\mathrm{exp},k}$, and incident photon fluence rate $\dot{\Phi}_q$, the model-predicted detector signal is 

\begin{equation} S_k^{\mathrm{pred}} (d,G) = t_{\mathrm{exp},k} A_{\mathrm{pix}} \dot{\Phi}_q \int f_q(E)\, T_j(E)\, R(E;d,G)\, \mathrm{d}E . \label{eq:signal} \end{equation}

\noindent

Here $S_k^{\mathrm{pred}}$ is expressed in ADU. The photon fluence rate after copper filtration is correspondingly \begin{equation} \dot{\Phi}_k^{\mathrm{trans}} = \dot{\Phi}_q \int f_q(E)T_j(E)\,\mathrm{d}E , \label{eq:filtered_fluence} \end{equation} which makes explicit how the number of photons incident on the detector is obtained for each filter condition. Air attenuation was not included separately, since it is already accounted for in the primary spectra $f_q(E)$.

\subsubsection{Response reconstruction}
\label{sec:fp_response}

The detector response was reconstructed by fitting the forward model of Eq.~\eqref{eq:signal} to the measured flat-panel signals for all calibration conditions $k$. The free parameters of the fit are the CsI:Tl effective scintillator thickness $d$ and the conversion factor $G$.

For each calibration condition, ten images were acquired with the selected
copper filter and ten corresponding open-beam flat-field images were acquired
without the filter. The images in each set were averaged before flat-field
correction and signal extraction to improve statistical precision.
\noindent
Before signal extraction, the images acquired with copper attenuation filters were corrected for pixel-dependent background and
gain variations using a background-corrected flat-field image. The corrected
signal at pixel $(x,y)$ was calculated as

\begin{equation}
I_{\mathrm{corr}}(x,y)
=
\frac{
I_{\mathrm{raw}}(x,y)-I_{\mathrm{dark}}(x,y)
}{
I_{\mathrm{FF}}(x,y)-I_{\mathrm{dark}}(x,y)
}
\,
\overline{
I_{\mathrm{FF}}-I_{\mathrm{dark}}
},
\label{eq:flat_field_correction}
\end{equation}

\noindent

where $I_{\mathrm{raw}}(x,y)$ is the measured raw image,
$I_{\mathrm{dark}}(x,y)$ is the detector background image, and
$I_{\mathrm{FF}}(x,y)$ is an image acquired under spatially uniform
irradiation. The overbar denotes the mean background-corrected flat-field
signal over the region of interest of the flat images. This normalization preserves the
mean signal level while correcting pixel-to-pixel gain variations. The
procedure follows the standard offset/gain flat-field correction used
for digital X-ray panel detectors~\cite{yu2013heel,seibert1998flat}

\begin{figure*}[htb!]
    \centering
    \includegraphics[width=1.0\textwidth]{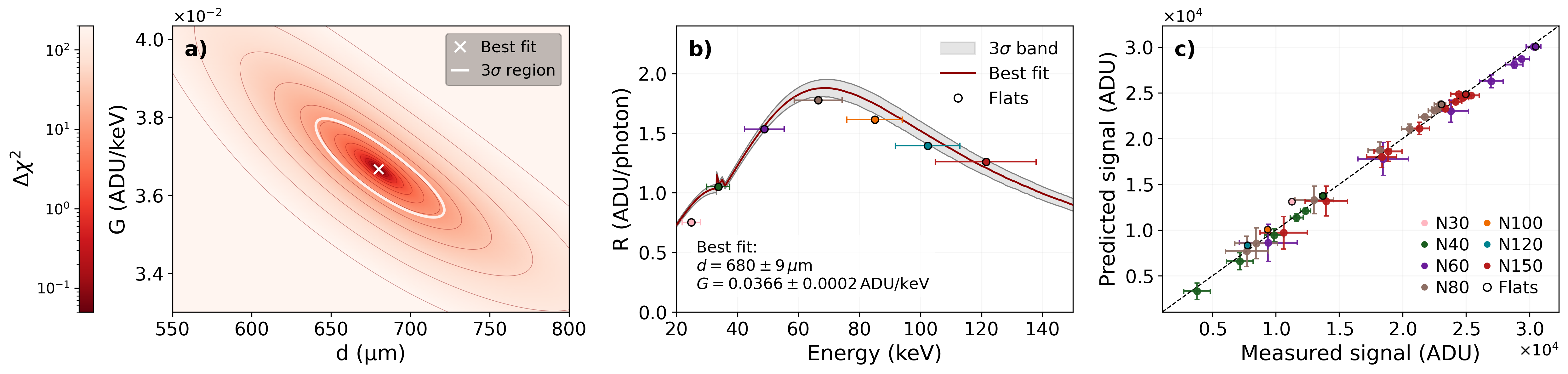}
    \caption{Summary of the flat-panel response-model reconstruction.
\textbf{(a)} Two-dimensional $\Delta\chi^2$ landscape in the $(d,G)$ parameter space. The white cross marks the best-fit parameter combination, and the white contour denotes the $3\sigma$ confidence region.
\textbf{(b)} Detector response $R(E;d,G)$ reconstructed using the forward model. The red curve shows the best-fit response, and the shaded region represents the propagated $3\sigma$ uncertainty band. Circular markers show effective response estimates derived from open-beam flat-field measurements by approximating each polychromatic calibration field as monoenergetic at its spectrum-weighted mean energy. These points are shown for qualitative comparison only; unlike the forward-model fit, they do not account for the full spectral shape of the incident X-ray field. The fitted values of $d$ and $G$ are reported in the panel.
\textbf{(c)} Comparison of the measured detector signals, $S_k^{\mathrm{meas}}$, with the corresponding model predictions, $S_k^{\mathrm{pred}}$, for all beam-quality and filter combinations included in the fit. The dashed line indicates the ideal parity relation, $S_k^{\mathrm{pred}} = S_k^{\mathrm{meas}}$.}  
    \label{fig:response}
\end{figure*}

The measured signal $S_k^{\mathrm{meas}}$ for each calibration condition $k=(q,j)$ was calculated as the mean pixel value (MPV) in analog-to-digital units (ADU) within regions of interest (ROI). For the copper-filter images, the ROIs were defined to cover the detector regions irradiated through the corresponding copper-filter sectors. For the unfiltered flat-field images, the ROI was calculated over the full active detector area. The standard deviation of the pixel values in ADU within the ROI was used as the nonuniformity contribution to signal error $\sigma_{S_k}$.

The best-fit values of $d$ and $G$ were obtained by minimizing the chi-square function
\begin{equation}
\chi^2(d,G)
=
\sum_k
\frac{
\left[
S_k^{\mathrm{meas}}-S_k^{\mathrm{pred}}(d,G)
\right]^2
}{
\sigma_{S_k}^2
},
\label{eq:chi2}
\end{equation}

\noindent
where $S_k^{\mathrm{meas}}$ and $S_k^{\mathrm{pred}}$ are the measured and model-predicted detector signals for calibration condition $k$, respectively, and $\sigma_{S_k}$ is the total uncertainty associated with the measured signal. The minimization was performed using the Levenberg--Marquardt nonlinear least-squares algorithm~\cite{levenberg1944method,marquardt1963algorithm}, with Eq.~\eqref{eq:chi2} as the objective function. This approach is well suited to the present problem because the detector model depends nonlinearly on the fit parameters, while the simultaneous use of all beam-quality/filter combinations provides sufficient constraint for a stable two-parameter fit.

Calibration conditions for which copper attenuation excessively reduced the signal-to-noise ratio were excluded from the fit. 
Besides the nonuniformity error described earlier, the signal uncertainty $\sigma_{S_k}$ includes contributions from frame-to-frame variability, estimated from repeated measurements, and from the uncertainty in the copper-filter transmission, obtained by propagating the filter-thickness uncertainty through Eq.~\eqref{eq:cu_transmission}~\cite{taylor2022introduction}.

The resulting $\Delta\chi^2$ landscape in the $(d,G)$ parameter space is shown in Fig.~\ref{fig:response}(a). A well-defined minimum is observed, indicating that the calibration dataset provides a unique and stable solution within the investigated parameter range. The elongated shape of the minimum reflects the expected partial correlation between effective scintillator thickness and conversion factor: increasing $d$ increases the fraction of photon energy absorbed in the scintillator, and this increase in signal can be partially compensated by a lower value of $G$.

The fit converged to an effective CsI:Tl scintillator thickness of $d = (680 \pm 9)~\mu\mathrm{m}$ and conversion factor of $G = (0.0366 \pm 0.0002)~\mathrm{ADU/keV}$. The reconstructed detector response is shown in Fig.~\ref{fig:response}(b), with the shaded band indicating the propagated $3\sigma$ uncertainty interval.
To evaluate the agreement between the fitted model and the experimental data, the predicted detector signals were compared directly with the measured signals for all calibration conditions. The comparison is shown in Fig.~\ref{fig:response}(c), where the data cluster closely around the identity line, indicating good agreement between the model and the measurements over the full range of investigated beam qualities and filter thicknesses.

\section{Calibration of the filter stack spectrometer}
\label{sec:calib-fss}
\subsubsection{Detector description} 
\label{sec:calo_desc}
The FSS used in this work consists of a stack of eight identical polyvinyltoluene-based scintillator tiles (EJ-200, Eljen Technology). EJ-200 has a density of 1.032 g/cm$^{3}$, a light yield of approximately 10,000 photons/MeV, and a peak emission wavelength of 425 nm. Each tile has a transverse cross-section of 2 cm $\times$ 2 cm and a thickness of 6 mm. The total length of the stack was optimized to improve the resolution of critical-energy measurements for synchrotron radiation with critical energies in the range of 10 to 50 keV.
\noindent

\begin{figure}[htb!]
    \centering
    \includegraphics[width=0.5\textwidth]{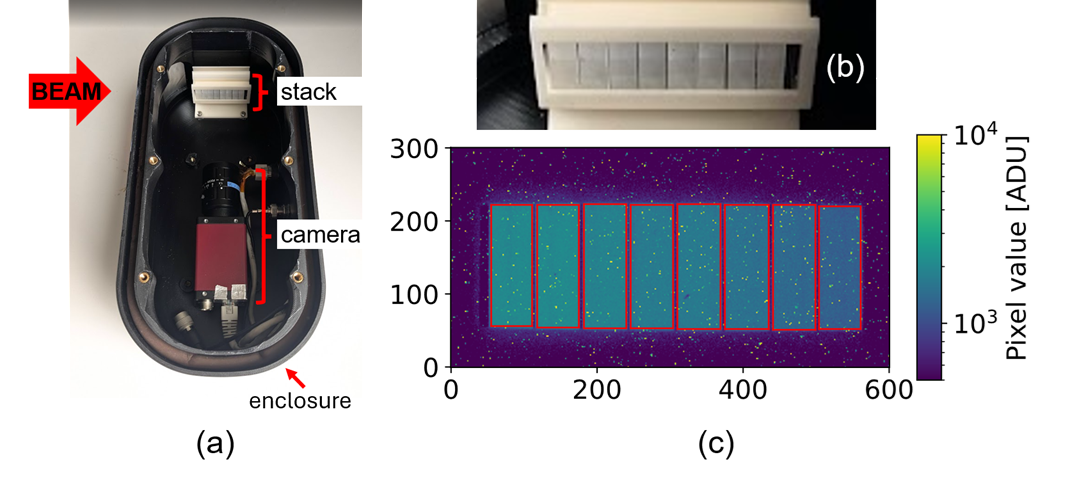}
    \caption{\textbf{(a)} Photograph of the FSS detector and \textbf{(b)} close-up view of the stack. \textbf{(c)} Typical camera frame with the position of the ROIs highlighted as red frames. The X-ray quality is N150 and the exposure time 10 s.}
    \label{fig:calo}
\end{figure}

The scintillator tiles were specially prepared to maximize light collection efficiency. First, all surfaces were polished except for one transverse face, which was roughened using 600-grit sandpaper. Next, all surfaces were coated with reflective titanium dioxide paint, with the exception of the face opposite the roughened surface. Finally, adjacent tiles were separated by 0.2 mm-thick black PVC spacers to prevent optical crosstalk between scintillator elements.
Scintillation light emitted by the tiles is recorded using the camera Manta G-325B from Allied Vision Technologies equipped with a lens 67709 from Edmund Optics. The camera gain was set to 1 and kept constant for all measurements. The uncoated face of each tile is oriented toward the lens featuring a focal length of 6 mm and a f-number of 1.4. The distance between the scintillator stack and the camera lens is 9 cm. The stack is mounted in a 3D-printed holder that ensures a good tile positioning. Both the scintillator stack and the camera are housed within a light-tight 3D-printed enclosure made of Polyethylene Terephthalate Glycol (PETG) plastic to eliminate ambient light contamination. To minimize the attenuation of incident X-rays while blocking ambient light, the enclosure includes an entrance window covered with a thin aluminum foil of thickness 20~{\textmu}m. A photograph of the detector assembly is shown in Fig.~\ref{fig:calo}(a).

\subsubsection{Irradiation with reference X-ray fields} 
\label{sec:calo_irr}
As mentioned previously, the FSS was irradiated using all X-ray qualities listed in Table~\ref{tab:xrayquality}. For each scintillator tile, a ROI corresponding to its visible area on the camera image was defined. The ROIs were defined to ensure approximately uniform light intensity across their areas, with the edges of the scintillation tiles excluded. These ROIs remained fixed throughout all measurements. The light output of tile $i$, denoted $L_i$, was defined as the MPV of all pixels within its ROI, expressed in ADU.

During irradiation, part of the camera signal originates from residual ambient light and from background radiation interacting directly with the camera sensor. To correct for these contributions, two background ROIs were defined above and below each signal ROI. The MPV of the corresponding background ROIs was subtracted from the signal MPV, yielding the corrected light output values $L_i$. A typical camera frame is shown in Fig.~\ref{fig:calo}(c).

For each X-ray quality, three images were acquired with exposure times of 1, 5, and 10 s. The light output rate, $\dot{L}_i$, expressed in ADU~s$^{-1}$, was determined from a linear fit of the corrected MPV values as a function of exposure time. This procedure minimizes the influence of possible offset effects associated with the camera hardware. The set of the eight light output rate values constitutes the detector's response vector.

\subsubsection{Simulation-driven response model} 
\label{sec:calo_sim}

The complete FSS system, including the scintillator stack and its enclosure, was modelled using the FLUKA Monte Carlo code. The simulation reproduces the experimental configuration by placing the photon source 100 cm from the entrance window of the enclosure. Photons are generated isotropically within a cone of 100 mrad divergence, corresponding to a solid angle of $\Omega_{\mathrm{sim}} = 0.00785$ sr. Their energy distribution is sampled from the spectra calculated with SpekCalc, the same as for the fluence calculations of Section~\ref{sec:calib-methology}, for each X-ray quality.

The medium surrounding the detector is modelled as vacuum because the SpekCalc spectra already account for attenuation in the 100 cm air path between the X-ray source and the detector. The energy deposited in scintillation tile $i$ is scored and denoted $E_{\mathrm{sim},i}$. FLUKA reports deposited energy per primary particle in units of $\mathrm{GeV} \, \gamma^{-1}$, where $\gamma$ denotes the primary photon.

The conversion from deposited energy per primary particle to energy deposition rate $\dot{E}_{\mathrm{sim},i}$ is given by

\begin{equation}
\begin{aligned}
\dot{E}_{\mathrm{sim},i}\,[\mathrm{GeV\,s^{-1}}] \; 
&= \; \dot{\Phi}\,[\gamma\,\mathrm{cm^{-2}\,s^{-1}}] \;
   \times \; E_{\mathrm{sim},i}\,[\mathrm{GeV}\,\gamma^{-1}] \\
&\quad \times \; \Omega_{\mathrm{sim}}\,[\mathrm{sr}]\;
 \times \;  r^{2}\,[\mathrm{cm^{2}\,sr^{-1}}]
\end{aligned}
\label{eq:calo-edep}
\end{equation}
\noindent
where $r^2$ represents the area per unit solid angle at the
source-to-detector separation $r$ (fixed at 100 cm in all measurement) 
\noindent
Owing to geometric point-of-view effects, the light collection efficiency $\varepsilon_\mathrm{POV}(z)$ decreases with increasing transverse distance from the optical axis of the imaging system. A correction for this effect was proposed in Eq. (9) of reference~\cite{LEFEBVRE2025107485}, yielding

\begin{equation}
\varepsilon_\mathrm{POV} (z) = \left(1+(z/L)^2\right)^{-\frac{3}{2}} \left( 1 - \frac{(z/L)^{2}}{n^2 \left( 1 + (z/L)^2 \right)} \right)^{-\frac{1}{2}},
\label{eq:pov}
\end{equation}
\noindent
where $L$ is the shortest distance between the lens and the scintillator stack, and $n$ is the refractive index of the scintillator material. The resulting reduction in measured light yield reaches at most 6\% for the outermost tiles.

The validity of Eq.~\ref{eq:pov} was verified experimentally by measuring the light output of individual tiles while irradiating them with a compact $^{137}$Cs source inside the enclosure. The simulated light yield is therefore expressed as $\dot{L}_{\mathrm{sim},i} = \varepsilon_\mathrm{POV} (z_i) \dot{E}_{\mathrm{sim},i}$ where $z_i$ is the transverse distance between the centre of tile $i$ and the camera optical axis.

\begin{figure}
    \centering
    \includegraphics[width=0.5\textwidth]{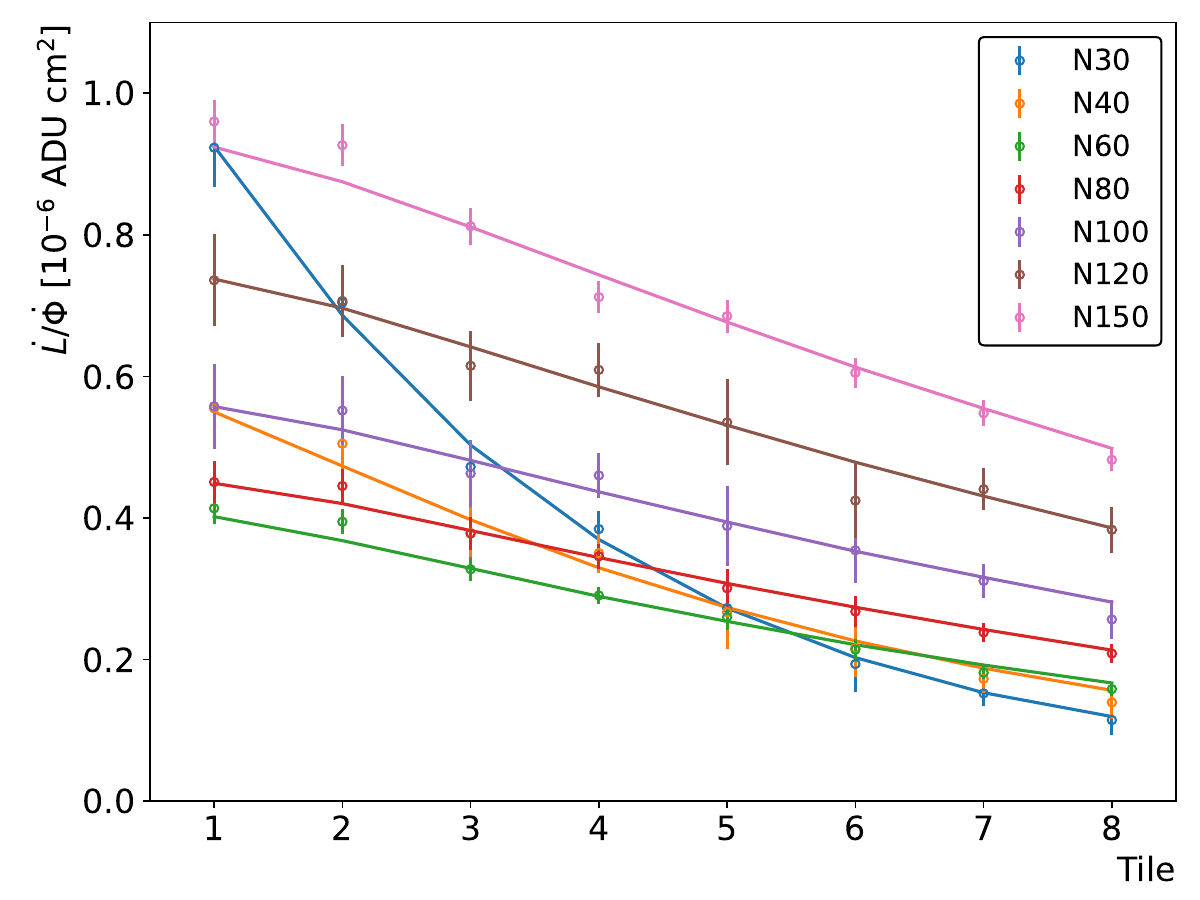}
    \caption{Measured FSS response vectors together with the corresponding best-fit simulation results. Symbols represent the measured light output rates, while solid lines show the scaled simulation predictions.}
    \label{fig:calo-results}
\end{figure}

The simulated response vectors can be converted into camera counts through $\dot{L}_{\mathrm{sim},i} = k  \dot{L}_{i}$ where $k$ is a universal calibration constant, assumed to be independent of both tile position and photon energy, with units of $\mathrm{ADU},\mathrm{keV}^{-1}$. This constant was determined independently for each X-ray quality by fitting the simulated response vector to the corresponding measured response vector. The optimal value was obtained by minimizing the weighted least-squares objective function
$\chi^2 (k) = \sum_i (\dot{L}_{i} - k \dot{L}_{\mathrm{sim},i})^2 / \sigma_i^2$
where $\sigma_i$ is the signal uncertainty, expressed in ADU, and evaluated using a procedure similar to that described in Section~\ref{sec:fp_response}.
The results are shown in Fig.~\ref{fig:calo-results} normalized to the fluence rate to improve the visualization while the best-fit calibration factors are reported in Table~\ref{tab:fssfit}.

\begin{table}[htb!]
\centering
\begin{tabular}{l|l|l}
X-ray quality & $k$ [$10^{-4}$ ADU keV$^{-1}$] & $\chi^2/\mathrm{NDF}$ \\
\hline
\hline
N30 &  1.23 $\pm$ 0.04 & 0.194 \\
N40 &  1.09 $\pm$ 0.04 & 0.360 \\
N60 &  1.00 $\pm$ 0.02 & 0.723 \\
N80 &  1.02 $\pm$ 0.02 & 0.203 \\
N100 & 1.02 $\pm$ 0.03 & 0.261 \\
N120 & 1.09 $\pm$ 0.03 & 0.269 \\
N150 & 1.13 $\pm$ 0.01 & 1.097 \\
\hline
\end{tabular} 
\caption{Values of the best-fit FSS calibration factors for each X-ray quality. The values of $\chi^2/\mathrm{NDF}$ indicating the fit quality are also reported.}
\label{tab:fssfit}
\end{table}

 The small values of $\chi^2/\mathrm{NDF}$ indicate a good fit quality for all X-ray qualities.
Averaging the results over all X-ray qualities yields $k = 1.09 \times 10^{-4} \, \mathrm{ADU}\,\mathrm{keV}^{-1}$ with a variation of 5\% between qualities. This variation is taken as the uncertainty of the calibration constant and likely reflects a combination of modeling inaccuracies and uncertainties in the assumed X-ray spectra. For example, the exact composition of the scintillation material is proprietary, and the additives incorporated into the PVT base are therefore unknown. Moreover, the light yields of the scintillation tiles are assumed the same while small variations are expected. A detailed investigation of these systematic effects is left for future work.

Despite this variation, the agreement between measurements and simulations validates the detector model within the stated uncertainty. The calibrated model therefore enables absolute X-ray fluence measurements, provided that a sufficient knowledge of the source spectrum is available.

\section{Conclusion}
\label{sec:conclusion}

We have calibrated and modelled two complementary scintillator-based detector systems for broadband X-ray diagnostics: the Varex Imaging XRD 0822 flat-panel detector and an in-house filter stack spectrometer developed at ELI Beamlines. Both systems were characterized using well-defined reference X-ray radiation fields, allowing their response to be evaluated under controlled and reproducible irradiation conditions. 

For the flat-panel detector, an energy-dependent response model was developed based on X-ray absorption in the CsI:Tl scintillator, with additional correction factors obtained from FLUKA simulations to account for incomplete energy deposition. The model was first checked using an effective-energy approximation of the reference beam qualities and was then reconstructed using the full spectral distributions together with copper filtration. The forward reconstruction yielded an effective CsI:Tl scintillator thickness of $680\,\mu$m and reproduced the measured detector signals across the investigated ISO-4037 N-series X-ray reference fields and filter configurations. 

The filter stack spectrometer was calibrated using the same reference X-ray fields. A FLUKA model of the scintillator stack was used to calculate the dose-deposition profile along the detector, including the experimental geometry and light-collection correction. Comparison between measured and simulated response vectors yielded a calibration constant of $k = 1.09 \times 10^{-4} \, \mathrm{ADU}\,\mathrm{keV}^{-1}$, with a 5\% variation between X-ray qualities. This agreement validates the simulation-driven response model within the stated uncertainty and enables absolute fluence measurements using the FSS, provided that the source spectrum is sufficiently constrained. 

Together, these calibrations establish a quantitative diagnostic basis for broadband LWFA X-ray measurements. The flat-panel detector provides large-area, spatially resolved measurements, while the filter stack spectrometer provides complementary sensitivity to the low-energy spectral component of the incident radiation, particularly below 50~keV, where the photon fluence emerging from the CsI
decreases and is modulated by absorption-edge effects. Their combined use is therefore well suited for the characterization of LWFA betatron sources, where broadband spectra and shot-to-shot fluctuations require detector systems with independently validated response models.

Additionally, both detectors are integral components of the diagnostic portfolio routinely employed during ELI-Beamlines user campaigns, particularly at the ELI Gammatron beamline. 
The exhaustive calibrations reported here contribute to reliable and well-characterized measurements and, in view of the use of these detectors in demanding laser-driven and accelerator-relevant experimental contexts, provide robust reference benchmarks that are directly applicable to user campaigns at ELI Gammatron, while also serving as valuable guidelines for a broad user community and for future experimental campaigns employing these detectors.


\acknowledgements

The authors gratefully acknowledge Jana Krchovová and Libor Judas from SÚRO for their support and assistance during the experimental measurements.

Antonia Morabito acknowledges the support and conducts her research under the Marie Skłodowska-Curie Actions - COFUND project, which is co-funded by the European Union (MERIT - Grant Agreement No. 101081195).

This work was supported by the Ministry of Education, Youth and Sports of the Czech Republic through the e-INFRA CZ (ID:90254). The authors acknowledge IT4Innovations National Supercomputing Center for granting access to computational resources through e-INFRA CZ.

Portions of this research were carried out at the ELI Beamlines Facility, a European user facility operated by the Extreme Light Infrastructure ERIC.

\section*{AUTHOR DECLARATIONS}

\subsection*{Conflict of Interest}

The authors have no conflicts to disclose.

\subsection*{Author Contributions}

Orsolya Morvai performed the formal analysis and visualization of the
flat-panel detector calibration and wrote the main part of the manuscript.
Benoit Lefebvre contributed substantially to drafting and revising the
manuscript and to the analysis and visualization of the filter-stack
spectrometer results. Marcel Lama\v{c} contributed to the calibration
methodology, scientific interpretation of the results, and review and
revision of the manuscript. Antonia Morabito contributed to the
conceptualization of the study, the framing of the broader relevance of
the calibration work, supervision of the project, and its development
toward publication. Uddhab Chaulagain secured funding for the calibration beamtime at S\'{U}RO, participated in scientific discussions,
and critically reviewed the manuscript. Petr Odstr\v{c}il,
Dominik \v{C}\'ap, Vojt\v{e}ch Janota, Alfred Haavaan Mishi,
Romain Caye, and Jaroslav Nejdl contributed to the experimental work,
participated in scientific discussions, and critically reviewed the
manuscript.

Orsolya Morvai and Benoit Lefebvre contributed equally to
this work. 

All authors reviewed and approved the final manuscript.

\section*{DATA AVAILABILITY}

The data that support the findings of this study are available from the
corresponding author upon reasonable request.

\section*{References}
\bibliographystyle{aipnum4-2}
\bibliography{aipsamp}

\end{document}